\documentstyle[twocolumn,aps,epsf]{revtex}
\begin{document}
\twocolumn[
\hsize\textwidth\columnwidth\hsize\csname@twocolumnfalse\endcsname
\draft

\title{Thermal Conductivity of superconducting (TMTSF)$_2$ClO$_4$:
 evidence for a nodeless gap}
\author{St\'{e}phane Belin and Kamran Behnia}
\address{Laboratoire de Physique des
Solides(CNRS), Universit\'{e} Paris-Sud, 91405 Orsay, France}

\date{June 2 1997}
\maketitle

\begin{abstract}
We report on the first measurements of thermal conductivity in the
superconducting state of (TMTSF)$_2$ClO$_4$.  The electronic
contribution to heat transport is found to decrease rapidly below T$_c$,  indicating 
the absence of low-energy electronic excitations. We argue that this result provides 
strong evidence for a nodeless superconducting gap function but does not exclude
a possible unconventional order parameter.
\end{abstract}
]

The (TMTSF)$_2$X family of quasi-one dimensional conductors (the Bechgaard
salts) are a well-known case of competition between superconducting and
Spin-Density-Wave ground states\cite{jerome}. At ambient pressure, most of
these extremely anisotropic compounds undergo a metal-insulator transition
at low temperatures and have a SDW fundamental state. Under moderate
pressure, the SDW instability is suppressed and replaced by a superconducting
transition at a critical temperature of the order of 1 K\cite{ishiguro}. One
exception to this scheme is (TMTSF)$_2$ClO$_4$ which is superconducting at
ambient pressure. The high-field properties of these compounds- including a
particular version of quantum Hall effect\cite{cooper} and commensurability
effects in the angular magnetoresistance\cite{osada}- have been intensely
studied during the past few years. However, in spite of early speculations
on a possible unconventional nature of superconductivity in this context\cite
{gorkov}, and contrary to the other families of exotic superconductors (i.e.
Heavy Fermions and cuprates), the superconducting state has been subject to
very few studies. The only attempt to explore the symmetry of
superconducting order parameter in a Bechgaard salt is reported by Takigawa,
Yasuoka and Saito\cite{takigawa}. These authors detected a T$^3$ temperature
dependence in the nuclear relaxation rate of proton in (TMTSF)$_2 $ClO$_4$
and concluded that the superconducting gap function should vanish along
lines on the Fermi Surface.

Thermal conductivity has proved to be a powerful probe of gap structure in a
number of unconventional superconductors. In the case of the heavy-fermion
superconductor UPt$_3$, thermal conductivity measurements constitute one
major source of our current knowledge on the angular distribution of nodes
in the gap function\cite{lussier,huxley}. In the case of YBa$_2$Cu$_3$O$%
_{6.9}$ , several convincing signatures of d-wave superconductivity have
been reported in a number of heat transport studies\cite{yu}. In the
Bechgaard salts, measurements of thermal conductivity have been restricted
to temperatures well above the superconducting instability\cite{choi}. In
this letter, we present the first study of heat transport in an organic
superconducting system. Our conclusion happens to be rather surprising as we
find strong evidence for a nodeless gap.

\begin{figure}[tbp]
\epsfxsize=8.5cm
$$\epsffile{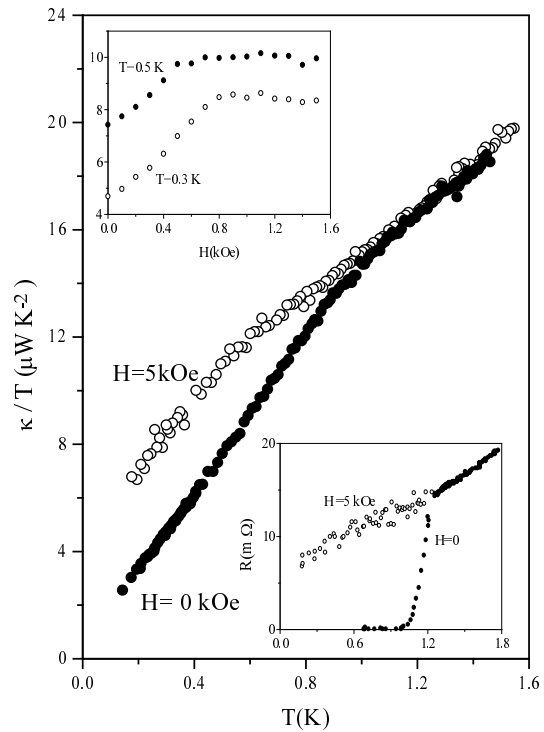}$$
\caption{Thermal conductivity divided by temperature for a relaxed sample of
of (TMTSF)$_2$ClO$_4$. Superconductivity is suppressed by applying a small
field along the c-axis. The lower insert presents the temperature dependence
of the electrical resistance for the same sample. The upper insert shows the
field dependence of $\frac \kappa T$ for two different temperatures.}
\label{fig1}
\end{figure}

The electrical and the thermal conductivities of four (TMTSF)$_2$ClO$_4$
single crystals- grown by standard electrochemical technique- were measured
with a conventional four probe method. Contacts were realized using silver
paint on evaporated gold. Due to the poor conductivity of these contacts ($%
\sim $ 1$\Omega $ ), the heat current passing through the sample was
carefully checked using an Au-Fe thermocouple connected in series with the
sample. The temperature gradient along the sample was measured with two RuO$%
_2$ thermometers which were thermally coupled to the contacts through gold
wires. To reduce the heat loss, small solenoids of 50 $\mu $m
superconducting Nb-Ti wires were made to hold each thermometer and measure
its resistance. To test our apparatus, we used it to measure thermal
conductivity of 20$\mu $m wires of metallic alloys (Al and Au-Fe) and found
a linear thermal conductivity in agreement with the Wiedemann-Franz (WF) law
and a Lorenz ratio very close to the Sommerfeld value (L$_0$= 2.45 10$^{-8}$ 
$\Omega $ m k$^{-2}$). All samples studied in this work showed jumps in
resistance due to appearance of cracks during the cooling process. This has
been regularly reported in transport studies of Bechgaard salts with
silver-paint contacts and makes the determination of the absolute value of
conductivity at low temperatures very difficult. However, the ratio of room
temperature resistance to residual (i.e. T $\rightarrow $0) resistance (RRR)
was found to be very different from one sample to another. For slowly cooled
samples (see below on the effect of cooling rate) this ratio was found to go
from 10 to 440. Here we present the results for the sample with a RRR of 440
(dimensions: 1.1 x 0.23 x 0.07 mm$^3$) which was most thoroughly studied.
But the same basic features were observed for the three other samples.

Fig.1 shows the temperature dependence of $\frac \kappa T$ and $\rho $ at
low temperatures. The superconducting transition leads to a sudden decrease
in $\kappa $ at T$\sim $ 1K which is coincident with the end of the
resistive transition. This kink in $\kappa $(T) disappears with the
destruction of superconductivity under a small magnetic field along the
c-axis. The ratio of thermal and electrical conductances at the onset of
superconductivity indicate that heat transport is dominated by phonons and
that the electronic contribution counts only for a small fraction of total
thermal conductivity for T 
\mbox{$>$}
1 K.

In general, the separation of lattice and quasi-particle components of
thermal transport in superconductors is not straightforward, as the
condensation of electrons in the superconducting state affects lattice
contribution due to electron-phonon coupling. To gain insight on the effect
of the superconducting instability on heat carriers, we plot in Fig. 2, $%
\frac{\Delta \kappa }T$, the difference between the two experimental curves
of $\frac \kappa T$ (at H=0 and H= 5 kOe), as a function of temperature. As
seen in the figure, upon the entry in the superconducting state, $\frac{%
\Delta \kappa }T$ increases steadily with decreasing temperature before
saturating at a temperature of about 0.4 K. This saturation has been
observed at the same temperature for all the samples studied in this work
and its value (3.7 $\pm $0.4 $\mu $W K$^{-2}$ in this sample) was found
every time to be close to $\frac{L_0}{R_0}$ (= 3.8 $\pm $0.5 $\mu $W K$^{-2}$
here) which is the expected maximum electronic contribution to heat transport 
according to the WF law.

According to recent theoretical results\cite{kane}, the separation of spin
and charge degrees of freedom in an interacting 1D electron gas can lead to
the violation of WF law and -in certain cases- to a divergence of Lorentz
number at zero temperature. Here, the correlation between the
zero-temperature extrapolations of normal state resistivity and the loss in
electronic thermal conductivity constitute the first confirmation of the WF
law in a quasi-one-dimensional conductor. This is not very surprising, since
below the temperature scale defined by the interplane coupling t$_c$
(estimated to be a few Kelvins), (TMTSF)$_2$ClO$_4$ is expected to behave as
an anisotropic 3D Fermi liquid. At finite temperatures, according to our
data thermal conductivity exceeds the limit imposed by the WF law. But this
additional thermal conductivity is within our experimental uncertainty on
the absolute value of the Lorentz ratio. Moreover, cracks which have a less
dramatic effect on thermal transport may cause a difference in the
geometric factors for thermal and electrical transport. Therefore, at this
stage, we will prudently remain within the boundaries of the WF law. Further
studies of thermal transport in the normal state under higher magnetic field
may elucidate this matter.

\begin{figure}[tbp]
\epsfxsize=8.5cm
$$\epsffile{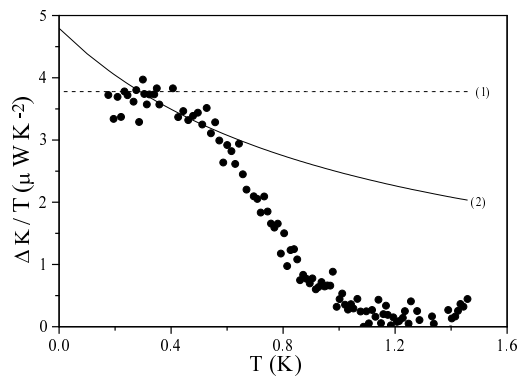}$$
\caption{The difference between the thermal conductivity of (TMTSF)$_2$ClO$_4
$ at H=5kOe and H=0kOe. Note the saturation at T 0.4 K. The horizontal line
(1) represents $\frac{L_0}{\rho _0}$. (2) schematizes a second scenario for
temperature dependence of thermal conductivity in the normal state (see
text).}
\label{fig2}
\end{figure}

It can be shown that the saturation in $\frac{\Delta \kappa }T$(T)
constitutes a strong argument in favor of the absence of low-energy
quasi-particle heat carriers. Neglecting the magnetoresistance (which is
very small at 5kOe as seen in the insert of Fig. 1), one can express this
difference as:

\[
\frac{\Delta \kappa }T(T)=(\frac{\kappa _e^n(T)}T-\frac{\kappa _e^s(T)}T)+(%
\frac{\kappa _{ph}^n(T)}T-\frac{\kappa _{ph}^s(T)}T)
\]
where subscripts ($e$, $ph$) stand for electronic and lattice components and
superscripts ($s$, $n$) refer to superconducting and normal states. Now, a
finite electron -phonon coupling would lead to an {\it increase} in the
lattice conductivity in the superconducting state so that $\kappa
_{ph}^n(T)\leq \kappa _{ph}^s(T)$ for the whole temperature range below T$_c$%
. This means that at any given temperature below T$_c$, $\frac{\Delta \kappa 
}T$ constitutes an lower limit to the difference between the electronic
conductivities of the normal and superconducting states. On the other hand,
at zero temperature this difference is given by $\frac{L_0}{\rho _0}$. These
two constraints will allow us to extract the temperature dependence of
normalized electronic thermal conductivity $\frac{\kappa _e^s}{\kappa _e^n}$
from $\frac{\Delta \kappa }T(T)$ and compare it to what is expected for
different gap structures. 
\begin{figure}[tbp]
\epsfxsize=8.5cm
$$\epsffile{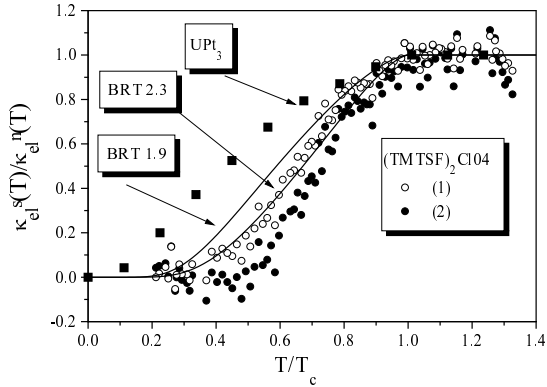}$$
\caption{Noramalized electronic thermal conductivity vs. normalized
temperature for (TMTSF)$_2$ClO$_4$ in two different scenarios (see text).
The results are compared with the predictions of BRT theory for two
different ratios of $\frac{\Delta (0)}{T_c}$ and with the published data on
UPt$_3$ for a heat current along the b-axis.}
\label{fig3}
\end{figure}

In Fig. 3., neglecting the effect of electronic condensation on lattice
conductivity, we consider two different scenarios for the temperature
dependence of thermal conductivity in the normal state. In the first
hypothesis, a constant $\frac{\kappa _e^n}T$ (equal to 
$\frac{L_0}{\rho _0})$, is assumed. In the second scheme, we suppose that 
the thermal conductivity in the normal state follows the behavior imposed 
by the temperature dependence of electrical resistivity and the WF law 
(curve (2) in Fig. 2). The latter scenario implies a difference of 27 
percent in the
geometric factor of the sample for electric and thermal transport. As seen
in Fig. 3, the normalized $\frac{\kappa _e^s}{\kappa _e^n}$ curve is not
very different for the two possible scenarios. It is instructive to compare
them with the data on UPt$_3$\cite{lussier}, the archetypal unconventional
superconductor. The decrease in the electronic thermal conductivity within
the entry in the superconducting state is much faster in (TMTSF)$_2$ClO$_4$.
At $\frac T{T_c}$ =0.4, for example, quasi-particle conductivity drops
virtually to zero in (TMTSF)$_2$ClO$_4$, but remains a sizeable (0.38)
fraction of normal state conductivity in UPt$_3$. Interestingly, our data
are much closer to the predictions of the conventional
Bardeen-Rickaysen-Tewordt theory\cite{brt}. The BRT function was computed
for different values of where $\frac{\Delta (0)}{T_c}$, where $\Delta (0)$
is the amplitude of the superconducting gap at zero temperature. The closest
fit was obtained for $\frac{\Delta (0)}{T_c}$ = 2.3. This strong-coupling
value shall be compared with 1.9 which is what is expected from size of the
jump in specific heat at T$_c$\cite{garoche}. However, due to the neglect of
electron-electron collisions in the BRT model, one can shall be cautious in
a quantitative comparison. Note that a finite electron-phonon coupling would
lead to an even sharper decrease in $\frac{\kappa _e^s}{\kappa _e^n}$ below T%
$_c$. Thus, in spite of several simplifications to obtain the plots of Fig.
3, our main conclusion is a direct consequence of the saturation presented
in Fig. 2 and remains quite robust:{\it \ there is no plausible way to
reconcile our data with a gap function vanishing on the Fermi surface.}

\begin{figure}[tbp]
\epsfxsize=8.5cm
$$\epsffile{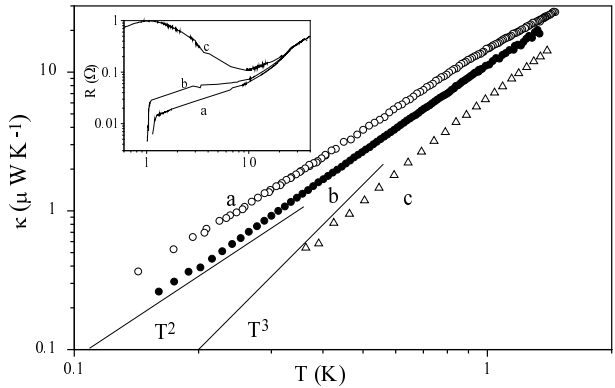}$$
\caption{Thermal conductivity vs. Temperature for different cooling rates:
(a) relaxed (0.6K/h) (b) intermediate (25K/h) and (c) quenched (60K/min).
Insert shows the temperature dependence of electrical resistance for the
three cases.}
\label{fig4}
\end{figure}
To gain further insight on lattice thermal conductivity, we studied the
effect of the cooling rate on low-temperature thermal conductivity. (TMTSF)$%
_2$ClO$_4$ passes through an anion-ordering transition at 24 K. The kinetics
of cooling around this temperature crucially influences the ground state.
The data reported and analyzed above were obtained on a relaxed
sample(cooling rate 0.6K/h) where the relative weakness of anion disorder
leads to a long mean-free-path both for phonons and electrons. The same
sample was warmed up to 40K in order to let the disorder fully develop and
then cooled down with different cooling rates. In this way, we studied an
intermediate state(25 K/h) and a quenched state($\sim $60K/min). As seen in
the insert of Fig.4, while the intermediate state still shows metallic
behavior and a superconducting instability - with a lower T$_c$ and a higher
residual resistivity-, the quenched state becomes insulating because of a
SDW transition at 9K. The change in the low-temperature thermal
conductivity, shown in Fig. 4, is remarkable. The thermal conductivity is
dramatically reduced reflecting the sensitivity of lattice conduction to
anion disorder.

One can estimate the phonon mean free path in different states using the
classical phonon gas equation $\kappa _{ph}$=$\frac 13$ c$_{ph}$ v$_s$ l$%
_{ph}$; where c$_{ph}$=$\beta $T$^3$ is the lattice specific heat ($\beta $%
=58$\mu $J/cm$^3$K$^4$\cite{brusetti}), v$_s$ is the velocity of sound (3
km/s is the reported value\cite{chaikin} for (TMTSF)$_2$PF$_6$ along the
a-axis ) and l$_{ph}$ is the mean free path. In this picture, the T$^{2.4}$
dependence of thermal conductivity in the quenched state, indicates that the
phonon mean-free-path increases very slowly with decreasing temperature and
can be estimated to be about 14$\mu $m at 400mK. This is one order of
magnitude smaller than the maximum allowed by the sample dimensions($\sim $
150 $\mu $m) and suggests that the disorderly domains of anion ordering in
the quenched state\cite{pouget} strongly scatter phonons. On the other hand,
in the relaxed state, where the thermal conductivity shows an essentially T$%
^2$ behavior below T$_{c,}$ the phonon mean-free path is estimated to be 100$%
\mu $m at 200mK. The ballistic regime and the associated cubic behavior in
thermal conductivity is only expected below 130 mK. Note that the effect of
cooling rate on heat transport confirms the smallness of electron-phonon
scattering. Indeed, in the quenched state, the phonons are much more
affected by anion disorder than the absence of electrons as scatterers.

The main outcome of this work is that the superconducting gap of (TMTSF)$_2$
ClO$_4$ has no nodes. This result is in contradiction with the conclusion of
the only other experimental investigation of gap structure in this system 
\cite{takigawa}. One shall note, however, that the temperature range where
the nuclear relaxation rate is reported to show a T$^3$ dependence is
limited to T 
\mbox{$>$}
$\frac{T_c}2$ \cite{takigawa}. This can not be considered as a convincing
evidence for nodes in gap, as even for conventional superconductors, the
exponential behavior is expected only at very low temperatures.

We would like to stress that a nodeless gap in this compound is not
necessarily associated with s-wave superconductivity. Enumerating possible
gap functions for a quasi-one-dimensional, Hasegawa and Fukuyama\cite
{hasegawa} showed that a pseudo-triplet(i.e. odd-parity) gap function with
no nodes is possible in the context of Bechgaard salts (see the t$_1$ state
in ref.\cite{hasegawa}). Indeed, here the order parameter can have opposite
signs for anti-parallel wave-vectors without vanishing anywhere on the Fermi
surface, due to the openness of the latter. The strongest argument in favor
of an odd-parity superconducting order parameter is the fascinating behavior
of the upper critical field. A recent study by Lee {\it et al.}\cite{lee} on
the sister compound (TMTSF)$_2$PF$_6$ has shown that when the magnetic field
is oriented in the most-conducting plane, the upper critical field exceeds
the Pauli limit. Insensitivity to this limit is naturally explained for
Cooper pairs in a triplet state\cite{gorkov}. More studies are required to
examine further this appealing possibility of an odd-parity gap function
without nodes. From a theoretical point of view, it is highly desirable to
establish all the possible gap functions for the Fermi surface of 
(TMTSF)$_2$ClO$_4$. Indeed, ref.\cite{hasegawa} neglects triclinic symmetry 
of the crystal and the low-temperature shape of the Fermi surface after
the opening of the anion-ordering gap.

In conclusion, we have measured the thermal conductivity of (TMTSF)$_2$ClO$_4
$ in the superconducting, metallic and insulating states. The results are
incompatible with the presence of nodes in the superconducting gap function.
Moreover, electrons are found to have little effect on heat conduction by
phonons.

We thank M. Ribault, L. Taillefer, for useful discussions, C. Lenoir for
providing us the samples and L. Bouvot for technical assistance.

\end{document}